\documentclass[12pt]{iopart}
\usepackage{graphicx}   
\usepackage{subfig}
\expandafter\let\csname equation*\endcsname\relax
\expandafter\let\csname endequation*\endcsname\relax
\usepackage{mathrsfs,amsmath}
\usepackage{siunitx} 
\usepackage[bottom=0.8in,left=1in,right=1in]{geometry} 
\usepackage{natbib}
\usepackage{array}
\usepackage{color}

\graphicspath{{figures/}}

\usepackage[us]{datetime}
\usepackage{fancyhdr}

\usepackage[utf8]{inputenc}
 
\usepackage[intoc]{nomencl}
\makenomenclature

\renewcommand\refname{}
\usepackage{dcolumn}
\usepackage{bm}
\usepackage{amsmath} 

\begin{document}


\title{Two-axis cavity optomechanical torque characterization of magnetic microstructures}

\author{G. Hajisalem$^{1,3}$, J.E. Losby$^{2,3}$, G. de Oliveira Luiz$^{1,3}$, V.T.K. Sauer$^{2,3}$, P.E. Barclay$^{1,3}$, and M.R. Freeman$^{2,3}$}

\address{ 
$^1$Department of Physics and Astronomy, University of Calgary, Calgary, Alberta T2N 1N4, Canada\\%
}
\address{ 
$^2$Department of Physics, University of Alberta, Edmonton, Alberta T6G 2E1, Canada\\%
}
\address{
$^3$Nanotechnology Research Centre, National Research Council
of Canada, Edmonton, Alberta T6G 2M9, Canada\\%
}

\ead{mark.freeman@ualberta.ca, pbarclay@ucalgary.ca}

\begin{abstract}
Significant new functionality is reported for torsion mechanical tools aimed at full magnetic characterizations of both spin statics and dynamics in micro- and nanostructures.  Specifically, two orthogonal torque directions are monitored and the results co-analyzed to separate magnetic moment and magnetic susceptibility contributions to torque, as is desired for characterization of anisotropic three-dimensional structures.  The approach is demonstrated through application to shape and microstructural disorder-induced magnetic anisotropies in lithographically patterned permalloy, and will have utility for the determination of important magnetic thin-film and multilayer properties including interface anisotropy and exchange bias.  The results reflect remarkable sensitivity of the out-of-plane magnetic torque to the nature of small edge domains perpendicular to the applied field direction, and also contain tantalizing indications of direct coupling to spin dynamics at the frequency of the mechanics.   
\end{abstract}
\maketitle 

\section*{Introduction}

Among the variety of hybrid condensed matter system investigations now facilitated by nanocavity optomechanics, spin-mechanical systems in uniform magnetic fields occupy an interesting niche requiring high sensitivity to pure torques.  Mechanical measurements of magnetic torque fundamentally probe combinations of net magnetic moment and magnetic susceptibility anisotropy \cite{Chikazumi1997}.  Thin-film structures dominate the applications of micro- and nanomechanical torque measurements \cite{Rigue2012}.  In most applications to magnetometry of thin-film magnetic structures \cite{Moreland2001, Moreland2003}, the net moment contribution dominates, owing to large shape anisotropy inhibiting magnetizability out-of-plane.  Typically, such measurements are limited to applications where the torsion axis is in the plane of the specimen.  In contrast, the net {\em perpendicular} magnetic torque is in-plane anisotropy dominated for generic planar specimens.  Incorporating additional capability to transduce torques perpendicular to the plane of the specimen would increase markedly the total information obtainable from torque sensing.  

Ideally, a mechanical device would be capable of transducing both in-plane and out-of-plane torque, yielding both sets of information about the sample.  Recent advances in nanomechanical displacement detection with optical nanocavities directly enable this preferred scenario.  Unlike conventional paddle interferometry, where the signal arises only from out-of-plane motion, the nanocavity can have first-order sensitivity to movement in more than one orthogonal coordinate.\footnote{Similar measurements could be implemented with a conventional device, using beam-clipping instead of interference to detect the in-plane twist.  While useful in some cases, the approach would yield both smaller signals, and higher noise (such as from beam-pointing instability and low-frequency mechanical vibration translating to optical intensity fluctuations).}  Traditionally, two-axis torque measurements on thin films have necessitated re-mounting a specimen in an orthogonal configuration, for a second series of measurements \cite{Collins1971}.  A two-axis vector torque sensor with piezoresistive readout was designed for torque measurements on cuprate superconductors, but not utilized extensively owing to a strong variability of performance with temperature \cite{Kohout2007a}.  In the context of sensitive nanoscale force measurements, atomic (vertical) force microscopy \cite{Binnig1986} quickly led to a complementary lateral force mode, first introduced in a study of atomic-scale friction \cite{Mate1987}.  This evolved into routine simultaneous measurements monitoring both flex and twist of the probe cantilever \cite{Meyer1990, Huang2004}.  

Previously, the high sensitivity of optical nanocavity torque readouts enabled versatile measurements with a single mechanical mode \cite{Wu2017, Kim2013, Yu2016}.  Operation under ambient conditions facilitated adjustment of the field geometry to vary the measurements between conventional net moment hysteresis and a unique setup - one where the same component of torque would additionally manifest off-diagonal susceptibility contributions arising from microstructural inhomogeneity in the sample.  The second torque mode, in contrast, allows the latter effects to be probed directly through the diagonal susceptibility, and independently of other torque contributions.  

\section*{Details of the Experiment}

Magnetic torques in constant (DC) fields -- as encountered in the operation of a mechanical magnetic compass -- arise from the variation of the total energy of the magnet as a function of the direction of the external applied magnetic field \cite{Cullity2008}.  Torques from micro- and nanomagnets are usually transduced in AC measurements using a small driving field applied perpendicular to the DC field direction and varying with time to pump a mechanical torsion resonance, which increases sensitivity when off-resonance measurements are limited by a fixed technical noise floor.  The AC signals are proportional to the gradient of torque versus magnetic field strength for the AC field component.  This admits a ready description in terms of vector cross-products of moment and field.  Assuming differential magnetic volume susceptibilities $\chi_i$ to account in linear response for the alternating moment generated in response to the RF driving field (and noting but neglecting off-diagonal terms from magnetic anisotropy that could couple if there were additional components of DC field \cite{Wu2017}), yields

\begin{equation} \tag{1a}
\tau_y^{\rm RF} = -m_x \mu_0 H_z^{\rm RF} + H_z^{\rm RF}\chi_z V \mu_0 H_x^{\rm DC} ,
\end{equation}
\begin{equation} \tag{1b}
\tau_z^{\rm RF} = +m_x \mu_0 H_y^{\rm RF} - H_y^{\rm RF}\chi_y V \mu_0 H_x^{\rm DC} .
\end{equation}

\noindent Previous in-plane RF torque studies of thin-film structures have relied heavily on the out-of-plane susceptibility term being negligible in comparison to the moment term \cite{Moreland2003, Davis2011}.  If out-of-plane torques are measured, in contrast, the two contributing torques will be of similar magnitude.  

The sensor structure for the present study is a silicon micromachined planar mechanical torsion resonator, read out by a photonic crystal nanocavity and integrated with a ferromagnetic sample of patterned thin-film permalloy.  Complete sensor details can be found in Ref.$\thinspace$\cite{Wu2017}.  The complete measurement device includes the tapered fibre coupling to the optical cavity\cite{Wu2014}, which for the present work has been optimized for transduction of the $\tau_z$ mode.  In retrospect the $z$-mode was present but very weak in the spectrograms of Ref.$\thinspace$\cite{Wu2017}, and had yet to be identified prior to work in Ref.$\thinspace$\cite{Hryciw2015} in which adjustment of the tapered fiber position within the near field of the nanocavity allowed tunable mechanical mode spectroscopy.   

\begin{figure}[ht] 
	\centering
	\includegraphics[scale=0.98] {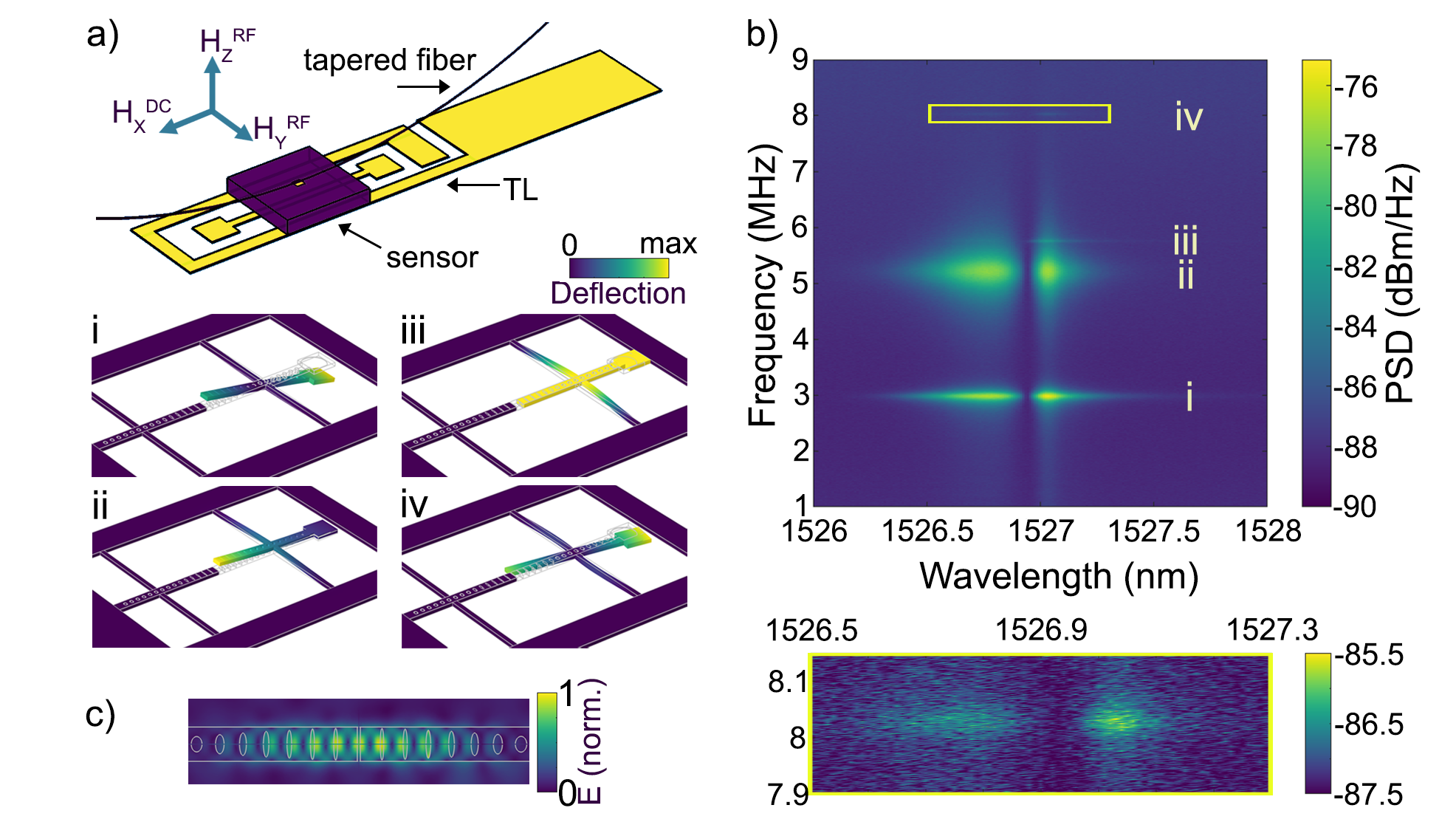}
	\caption{a) Schematic of the experimental setup, showing the tapered optical fibre used to evanescently couple light into and out of the nanocavity optomechanical device (location indicated by a yellow square on top of a purple substrate). The planar transmission lines (yellow traces, TL) are used to generate both $y$- and $z$-direction RF driving fields.  Zoomed-in detail of the photonic crystal nanocavity torque sensor (the thin-film permalloy island is deposited onto the paddle on the right-hand side) is shown in panels i - iv, comprising the first four mechanical eigenmodes of the device as determined by finite-element modelling \cite{COMSOL}.  b) Spectrograph of Brownian position noise across the frequency range of the first four modes, as a function of laser probe wavelength when it is swept across the cavity resonance wavelength and read out by tapered optical fibre coupled evanescently to the nanocavity.  A zoom-in of the $z$-torque (iv) mode signal is shown below. c) Simulated normalized electric field magnitude within the nanocavity, showing high field concentration in the gap between the photonic crystal Bragg mirrors.} 
	\label{Fig1}
\end{figure}

Fig.$\thinspace$1a summarizes the experimental setup and shows a set of mechanical eigenmodes, each illustrated through an exaggerated rendering of an instant of maximum displacement in the mode as determined by finite-element simulation \cite{COMSOL}.   The $z$-torque mode of the torsion pendulum, as gauged by a figure of merit capturing its relative sensitivities to twisting and flexing (see Supplementary Material), is in fact superior to the $y$-torque mode that has been the sole object of focus previously.  Figure 1b shows a 2D plot of measured optical noise spectra obtained from the photodetected signal of a laser transmitted through the fibre taper for varying wavelength within the optical cavity mode's Lorentzian lineshape, as in Ref.$\thinspace$\cite{Wu2017}. This noise spectrum captures the optomechanical transduction of the nanocavity's Brownian position fluctuations in air at room temperature via their interaction with the localized nanocavity mode, whose field profile is shown in Fig.$\thinspace$1c. Maximum transduction occurs near the points of maximum slope of the cavity mode lineshape, indicative of predominantly dispersive optomechanical coupling in the device \cite{Wu2014}. All four mechanical modes from Figure 1a are observed.  Notably, the in-plane translation of the paddle in the two higher frequency modes presents a smaller area for momentum-exchanging collisions with air molecules and results  in significantly less viscous damping \cite{Svitelskiy2009} and hence larger mechanical quality factors, $Q$.  With the structure offering excellent coupling to pure $y$- and $z$-torques respectively through the 2.99 and 8.02$\thinspace$MHz modes, the optical nanocavity enables the advances in magnetic measurement reported below. Previous studies focused on the 2.99$\thinspace$MHz mode (mode i), however the nanocavity is uniquely suited to detection of the corresponding displacement for both of the torsional modes at the photonic crystal mirror end of the sample support nanobeam where it interacts with the confined nanocavity optical mode.  

Differences in amplitude of the transduced signal of the four modes' Brownian motion is in part related to the different optomechanical coupling coefficients characterizing their interaction with the nanocavity optical mode. In particular, modes with vertical displacement (i and ii) have been found to experience significant optomechanical coupling owing to the presence of the optical fibre taper which breaks the vertical symmetry of the device, inducing both dissipative and dispersive optomechanical coupling whose relative strength depends sensitively on the precise fiber position \cite{Hryciw2015}. Mode iii is expected to have significant intrinsic dispersive coupling, as its motion directly modulates the effective optical cavity length. In contrast, mode iv has nominally zero dispersive optomechanical coupling due to the odd symmetry of its motion relative to the optical cavity mode intensity profile. However, fabrication imperfections combined with dissipative optomechanical coupling contributions result in the observed transduced signal. In addition, asymmetric lateral positioning of the fibre taper to the side of the nanobeam will induce dispersive optomechanical coupling for this $z$ torque mode, in a manner analogous to the broken vertical symmetry induced coupling from the fibre taper for modes i and ii. This coupling is harnessed in the measurements reported below to probe $z$ torques; in combination with the relatively high mechanical $Q \sim 138$ this mode provides comparable torque sensitivity (9$\thinspace$zNm/$\sqrt{\rm Hz}$) as the $y$ mode (17$\thinspace$zNm/$\sqrt{\rm Hz}$), as discussed in Supplementary Material - S2. 


\section*{Two-axis torque study of magnetic hysteresis: observations, and analysis of the coupled system of equations}

Figure 2 shows representative, averaged (9 sweeps) hysteresis curves for a cycling of in-plane DC field along $\vec{x}$ from 46$\thinspace$kA/m to zero and back, for the 2.99$\thinspace$MHz, $y$-torque conventional mode (panel a), and the 8.02$\thinspace$MHz, $z$-torque modes (panel b).  The $y$-torque is dominated by the net moment of the permalloy structure and evolves, as we have reported earlier \cite{Wu2017}, from a quasi-uniformly magnetized configuration at high field (characterized by uniform internal magnetization capped by small closure domains along the left and right edges), to a double vortex ground state at zero field.  The measured $\tau_y$ is proportional to the net moment along $x$, with a very small correction from a low $z$-susceptibility. 

\begin{figure}[ht] 
	\centering
	\includegraphics[scale=1] {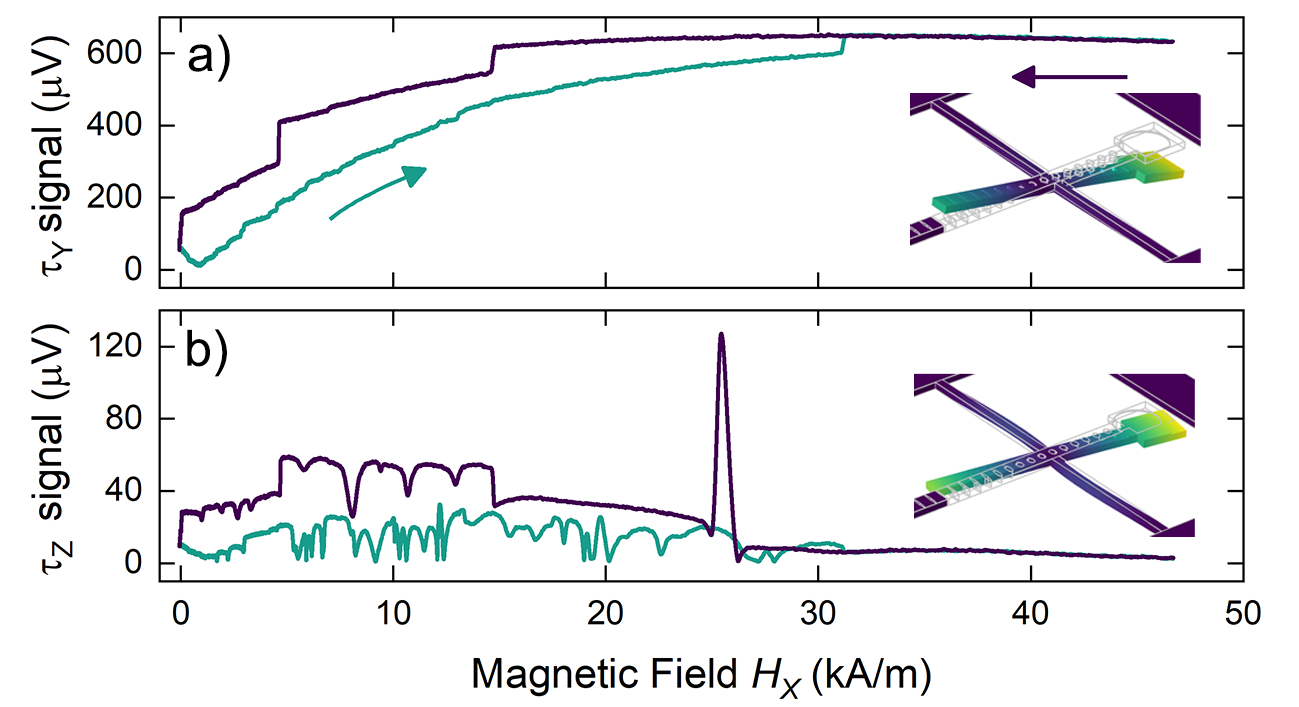}
	\caption{Experimental measurement of magnetic hysteresis of the permalloy structure for a cycling of in-plane DC field along $x$-direction (averaged 9 sweeps). Magnetization response for a) the 2.99$\thinspace$MHz $y$-torque mode driven with an RMS amplitude of 122$\thinspace$A/m, and b) the 8.02$\thinspace$MHz $z$-torque mode driven at 130$\thinspace$A/m RMS. Insets show the corresponding simulated mechanical modes for 2.99$\thinspace$MHz and 8.02$\thinspace$MHz.
	\label{Fig2}}
\end{figure}

The structure can be thought of as two connected shapes each having a single vortex ground state, and for which their collective ground state develop vortices of opposite chirality such that the connecting region is magnetized without any intervening domain wall.  In the upper branch of the hysteresis (field sweep down, purple), the first large step-down in $y$-torque, at 15$\thinspace$kA/m, is a vortex nucleation at the curved, left edge of the structure.  The next irreversible, large steps down in net $x$-moment corresponds to the jump of a domain wall into the right-hand side.  Between these two steps, the vortex texture is highly asymmetric and connected to a quasi-uniform magnetization on the right-hand side.  As a consequence, the core displacement susceptibility (as a function of $H_x$) is anomalous through this field range, with the core moving along $x$, at right angles to its normal motion.  The third step-down in moment occurs when a second vortex nucleates on the right side.  In ramping the bias back from zero to high field the double vortex persists to $H_x^{DC} \approx 13\thinspace$kA/m, whereupon the core in the stem (right-hand side) annihilates and after which the structure remains in an asymmetric single vortex state until $H_x^{DC} \approx 32\thinspace$kA/m.  Barkhausen steps reflecting spin texture pinning at microstructural disorder (including edge roughness) become visible in the $y$-torque hysteresis commencing with the nucleation at 15$\thinspace$kA/m.  They become more numerous (more features per unit $H_x$, on average) in the vortex states where the cores behave as nanoscopic probes smaller than the permalloy grains, exploring the internal disorder \cite{Chen2012,Burgess2013}.  

The corresponding $\tau_z$ hysteresis gives a fully complementary perspective on the equilibrium magnetization of the structure.  Whereas the $y$-torque signal reflects the net moment on account of the $z$-susceptibility being very small, the two contributions to the $z$-torque, in comparison, have the same order of magnitude since the in-plane shape anisotropy is much smaller than its out-of-plane counterpart.  The net $z$-torque being non-zero can be thought of as a result from the unbalancing of those two terms.  At high $H_x^{\rm DC}$, the measured $\tau_z$ is nearly zero.  This means that, for large enough DC fields applied along $\hat{x}$, the total magnetic energy remains roughly constant over small changes of the in-plane field direction, and the RF torque vanishes.  In terms of the expression for net $z$-magnetic torque (Eqn.$\thinspace$2b), in the high $H_x^{\rm DC}$ field regime the first term is constant ($m_x$ saturated) while the second term also becomes field-independent at large enough field, on account of $\chi_y$ being inversely proportional to $H_x$. This is expected in the $H_y \ll H_x$, small angle limit and for uniform magnetization nearly parallel to the field direction (see Supplementary Material - S4). 

In early macroscopic torque studies of magnetocrystalline anisotropies (measurements performed on monocrystalline disks), different scaling behaviours were observed for the decreasing torque at high field, and discussed with reference to models accounting for the role of so-called edge or closure domains \cite{Schlechtweg1936, Tarasov1939, Kouvel1957}.  Saturating the structure {\em completely} to eliminate every vestige of magnetic moment tangential to the sample edges in the $y$-direction would require infinite $H_x$.  Micromagnetic simulations of $z$-torque for rectangular permalloy islands indicate that `butterfly' closure domains (characterized by the edge breaking into two segments magnetized oppositely along $y$) yield smaller out-of-plane torques than are produced by $H_y$ fields applied to `S-state' spin textures (single segment edge domains, and oriented in the same direction along both $y$-edges).  Simulations performed for the measured sample shape show it avoiding the S-state closure configuration.  


As the in-plane field is lowered, non-uniform spin textures arise through the interior of the structure and these magnetize anisotropically, leading to dramatic variations of net $z$-torque.  Looking first at the high-to-low field branch of the hysteresis, as $H_x$ drops the net $m_x$ initially remains higher than it would be if described, for example, by a constant susceptibility to saturation, and the net $\tau_z$ grows with decreasing field.  At low bias fields, the hysteresis measured with $z$-torque superficially appears to be uninterpretably fluctuating.  In fact it is rich with information, rendered with superb signal-to-noise ratios despite the small magnitude of torques involved, thanks to the optical nanocavity readout.  Neglecting for a moment the very large fluctuations in $z$-torque, the characteristic average torque as a function of $H_x$ corresponds to what one could expect to observe for a pristine, monocrystalline structure of the same shape, reflecting the overall in-plane shape anisotropy (and discussed further in relation to Fig.$\thinspace$4b.  The large fluctuations, in contrast, capture local anisotropies in the displacement susceptibility of the high energy density regions of the spin texture (domain walls, vortex cores) as they explore the internal microstructure.  Once again, these features are most populous in the double vortex state where the two cores explore independent regions of disorder.  For quasi-static response (in equilibrium with the total applied field, which includes the low radio frequency components), the core motions are expected to be largely uncorrelated.  In the opposite vortex chiralities arrangement, the cores are not linked by a high energy density interconnect such as a N\'eel domain wall, which would introduce a significant energetic component dependent on relative position.   

Minor hysteresis loops spanning a sub-interval of the full field range are presented in Fig.$\thinspace$3, showing the results of sweeping the field between $H_x=0$ and 25$\thinspace$kA/m.  This cycles the structure between its double and single vortex states, by restricting the maximum applied field to less than the annihilation field for the left-hand side vortex.  Intercomparing the loops measured using the orthogonal torsion modes, features (peaks or dips) in the $z$-torque generally outnumber the Barkhausen steps in the $y$-torque.  These characteristics are correlated in that there will be a peak or dip in $\tau_z$ at a Barkhausen step, but there can be additional nonmonotonic evolution of $\tau_z$ between the Barkhausen step field positions.  Furthermore there is no correlation between the sizes of Barkhausen steps and $\tau_z$ features.  A zoom-in scan of the lowest field region is included in the Supplementary Material - S5. 

\begin{figure}[ht] 
	\centering
	\includegraphics[scale=1] {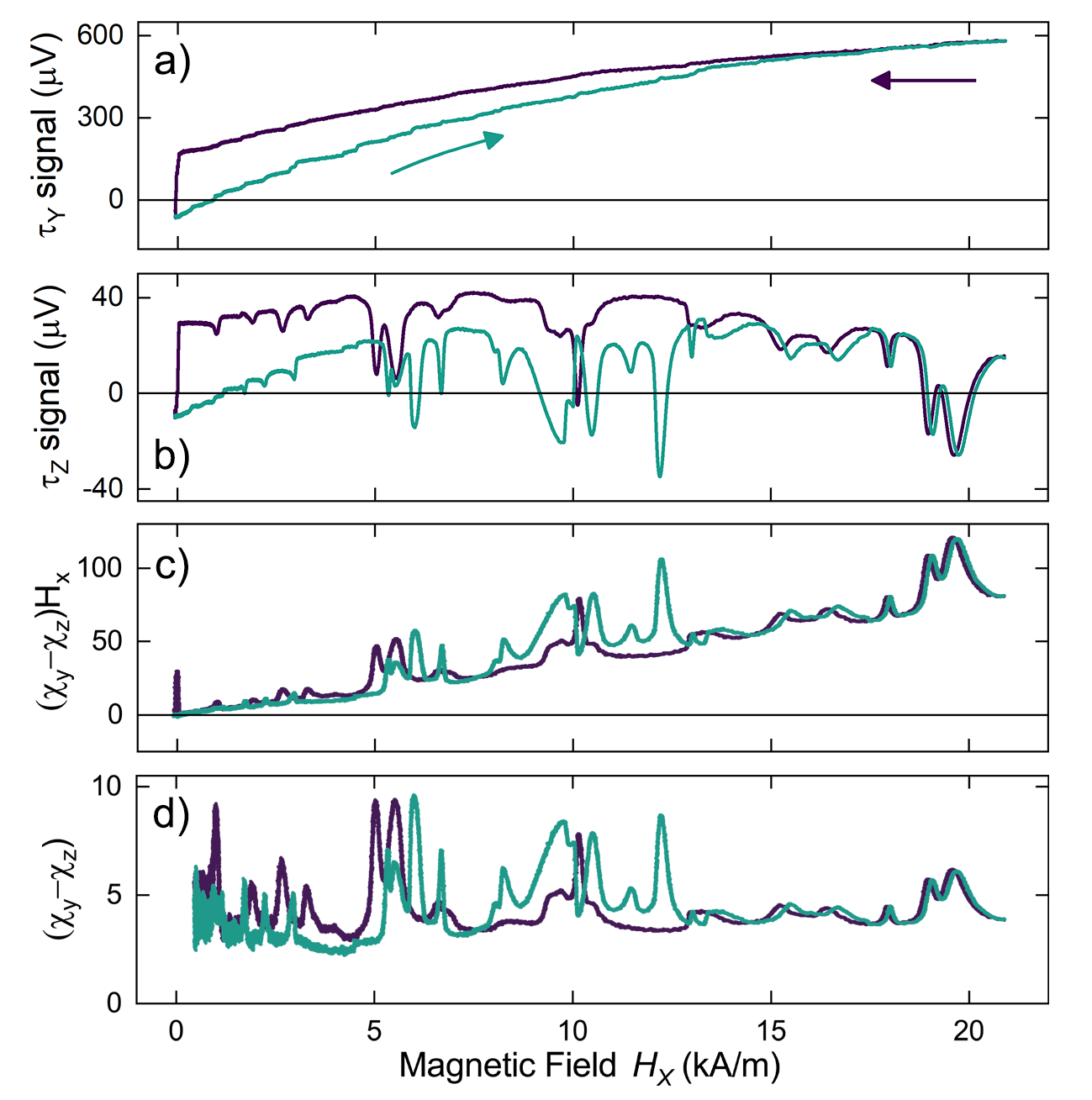}
	\caption{Experimental measurement of a minor hysteresis loop sweeping $H_x$ over a sub-interval (0 to 25$\thinspace$kA/m) of the full field range (averaged over 10 sweeps) for the 2.99$\thinspace$MHz (a) and 8.02$\thinspace$MHz modes (b). (c) Deduced differential magnetic susceptibility times field torque term calculated from the data in (a) and (b). (d) Differential magnetic susceptibility after dividing through the result of c) by the applied field strength, $H_x$.  The analysis leaves the results in c) and d) on the same scale as the $\tau_z$ signal in b): $\mu$V in c), $\mu$V/(kA/m) in d).  
	\label{Fig3}}
\end{figure}

Concentrating on the observations obtained from vortex states enables some insights into the basic processes giving rise to the strong modulations of $z$-torque.  As the in-plane magnetic field orientation is changed, the work done by magnetic torque on the individual moments integrates over the volume of the sample to yield a change in total magnetic energy.  This change in magnetic energy subdivides into changes in three mechanistic contributions, the Zeeman, demagnetization, and exchange energies (in the present experiment; in general there can be more, for example from interface anisotropy or exchange bias).  The breakdown of the energetic contributions to the internal torques provides further insight into why the $z$-torque is much more sensitive than the $y$-torque to disorder within the structure.  The energy landscape for field rotation out-of-plane (corresponding to in-plane torque) is dominated by the shape anisotropy of the thin-film structure.  A small out-of-plane moment develops to supply the internal counter-torque via demagnetization and Zeeman energies, but the spin configuration and associated exchange energy changes very little.  The out-of-plane field perturbation senses the net result of registration of the spin texture within the disorder potential, but not the `stiffness' of the registration against perturbations that attempt to translate the high energy density regions of the texture \cite{Burgess2014}.  

The out-of-plane magnetic torque, by contrast, has first-order dependence on the response of the texture to the in-plane perturbing field, including how the detailed registration to the static disorder makes it overall harder or easier to rotate than it would be on the flat energy surface of a pristine structure.  In this case, the internal counter-torque develops through a flexing of the spin arrangement that significantly engages the `exchange spring', in addition to changing the demagnetization energy.  Both effects are concentrated where the energy densities are highest — in vortex cores, and domain walls.   

Previous in-plane RF torque studies of thin-film structures have relied heavily on the out-of-plane susceptibility term being negligible in comparison to the moment term.  If out-of-plane torques are measured, in contrast, the two contributing torques will be of similar magnitude. 
The paired measurements of hysteresis through $\tau_z$ and $\tau_y$ can be analyzed in combination to extract the non-negligible in-plane differential susceptibility perpendicular to the DC field direction, and governing the net out-of-plane torque.  When the torque is hysteretic at or near $H_x =0$, a simple analysis can extract the quantity $(\frac{\partial m_y}{\partial H_y} - \frac{\partial m_z}{\partial H_z})H_x \equiv (\chi_y - \chi_z)H_x$ as a function of field without invoking any assumptions (see Supplementary Material, S3, for a full description and benchmarking of the procedure).  The result of this analysis as applied to the data of Fig.$\thinspace$3a,b) is shown Fig.$\thinspace$3c,d).  The simulation shows that for the present structure, $\frac{\partial m_z}{\partial H_z} \approx 0.04\frac{\partial m_y}{\partial H_y}$, independent of $H_x$.
Complete characterization of three-dimensional structures will require in addition measurements for a third torque axis, in order to resolve all components of the full system of equations.

\section*{Micromagnetic simulations of two-axis torque}

To elucidate basic features of the experiment, micromagnetic simulations are conducted to estimate the torques as a function of $H_x$ for two separate model systems: one to capture the overall effect of the shape of the permalloy structure on the torque hysteresis curves, and the other to explore how the torques are affected by the microstructural disorder within the film.  In order to adjust the relaxation built-in to the simulation script such that the output is representative of equilibrium magnetic behaviour, at each setting of $H_x$ a small cycling of $H_y$ between $\pm 80\thinspace$A/m and back to zero is computed to obtain the parameters required for estimating the $z$-torque using Eqn.$\thinspace$2b.  A linear fit to the loop output yields the slope of $z$-torque versus $H_y$.  The uncertainty of the fit correlates with the apparent noise in the plot of simulated $\tau_z$.  For the outlier points at spin texture transitions, the equilibration times allowed within the simulation were insufficient, and the linear fits were not meaningful.  A similar loop with fewer $H_z$ points is used for $\tau_y$, on account of the $H_z$ steps representing much smaller perturbations to the texture and the corresponding relaxation being faster.  All simulations are performed using the open source, GPU-accelerated mumax$^3$ platform \cite{Vansteenkiste2014, Leliaert2017}.   

The first model simulation, for shape effects, is summarized in Fig.$\thinspace$4a,b (the mask shape for the simulation is shown in the inset to Fig.$\thinspace$4a).  Most of the main features of the two-axis torque measurements are reproduced, with the exception of the very large peak in $\tau_z$ seen in Fig.$\thinspace$2 at $H_x = 26\thinspace$kA/m (a nonequilibrium effect, discussed later).  In addition, the left-side vortex does not annihilate when the field sweeps back up over this range, so some higher field hysteresis remains that isn't found in the experiment.  The simulations are not expected to accurately reproduce the transition fields, as vortex nucleation and annihilation is influenced by edge roughness and by thermal activation.  An animation showing the evolution of the spin texture throughout the simulated loop is included in the Supplementary Material, S6. In the animation the field is further increased until the vortex annihilates.  

For the second model simulation, microstructural disorder is imposed on a thin-film permalloy square in the form of a Voronoi tiling.  The tiling mimics surface roughness by populating the regions at random with two values of saturation magnetization differing by 5\%, and the physical effect of granularity (polycrystallinity) is included by reducing the exchange coupling between regions by 10\%.  The simulation is initialized in the so-called Landau state (the vortex state in a square includes 90$^{\circ}$ N\'eel walls along the diagonals, running from the core to the corners), at $H_x$=0.   

\begin{figure}[ht] 
	\centering
	\includegraphics[scale=1] {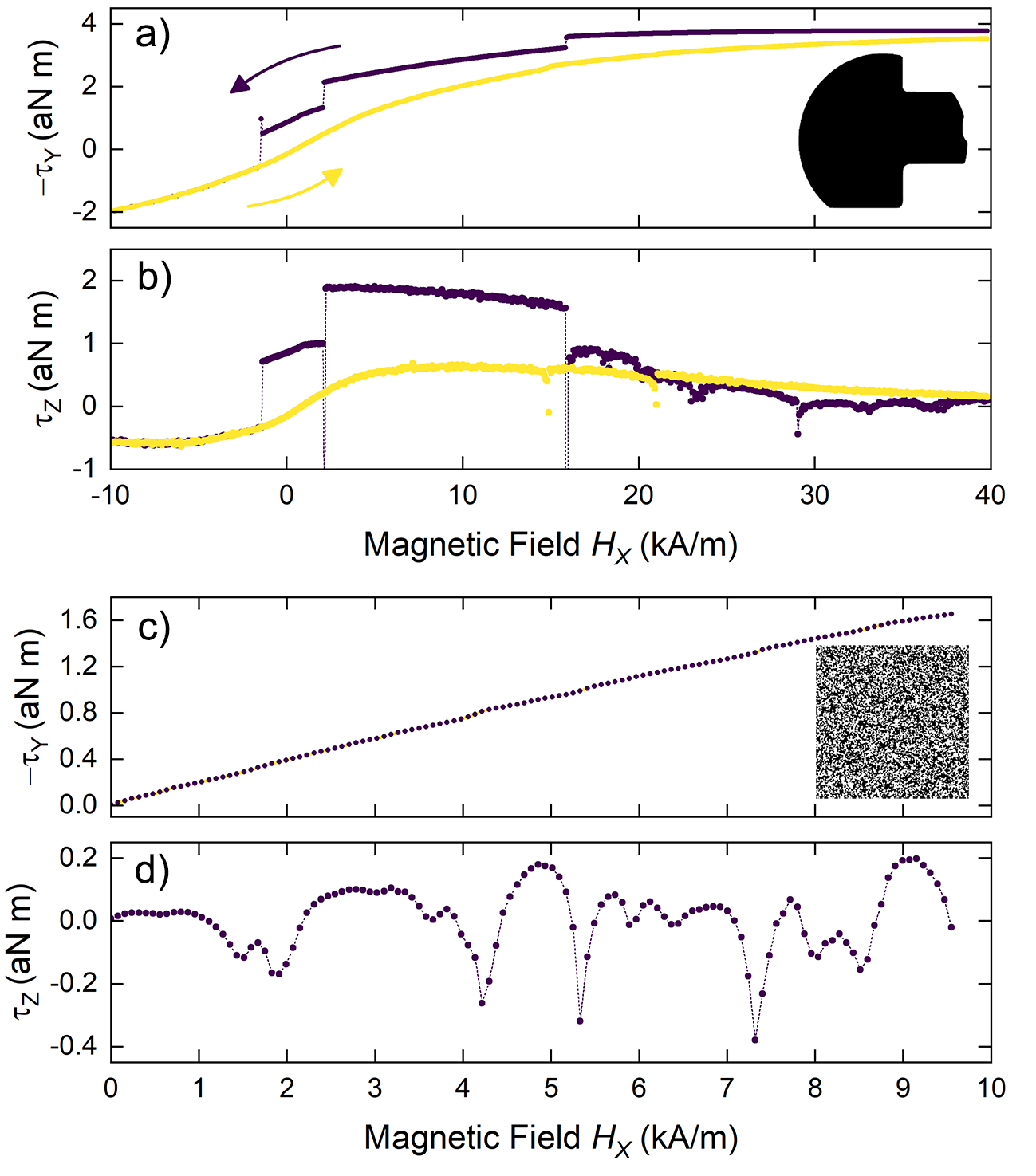}
	\caption{Micromagnetic simulation of the $y$- and $z$-torques generated in response to 80$\thinspace$A/m driving fields.  In order to parse the new information coming from the $z$-torque measurement, separate simulations examine: a,b), the effect of the sample shape (thumbnail inset in a); and c,d), the effect of microstructural disorder (inset thumbnail in c) shows the random tiling used to add the equivalent of 5\% surface roughness and a 10\% reduction of exchange coupling between grains).} 
	\label{Fig4}
\end{figure}

Fig.$\thinspace$4c,d demonstrates that essential qualitative features of the microstructural disorder on the measurements are also replicated via this procedure.  An average Voronoi region size of 10$\thinspace$nm is found to reproduce a rough approximation of the number of features observed in the experiment (the tiling pattern is shown in the inset to Fig.$\thinspace$4c.  An animation of the exchange energy density in the square is included in the Supplementary Material- S6, for visualization of the interaction of the spin texture with the disorder.  The animation reveals strong variations of the vortex core motion induced by $H_y$, dependent on whether the disorder potential locally favours or impedes core movement, from an energetic perspective.  In addition, the simulation reveals substantial modifications of the $y$-magnetic susceptibility stemming from particulars of how the 90$^{\circ}$ N\'eel walls conform to the disorder (the walls have a substantially lower energy density than the core itself, but occupy a much larger volume).  The role of the walls is confirmed in a companion simulation on a disk, where they are absent.  The domain wall mechanism found in simulation is consistent with the experimental finding in Fig.$\thinspace$3 of an increase in the number and magnitude of strong spikes between 5$\thinspace$kA/m and 13$\thinspace$kA/m on the field-sweep up, where the sample is expected to host an asymmetric Landau state on the right-hand side.

To reiterate, the simulations at this stage concentrate on equilibrium behaviour.  With more GPU power it will become possible to examine slow dynamics at the mechanical frequency, such as thermally-activated hopping events that can synchronize with the $H_y^{\rm RF}$ drive in a form of stochastic resonance to yield larger displacements of the core and correspondingly more dramatic features in the $z$-torque.


\section*{Out-of-plane quadrature RF torque observations}

The largest magnitude peaks in $z$-torque observed in this study occur outside of the vortex textures, in the hysteresis branch descending out of the quasi-uniform state.  To glean additional experimental insight into the nature of the features, it is useful to repeat the measurements for a range of small, constant transverse bias fields ($H_y^{DC}$) applied to the sample.  For the case of the magnetic vortex, this approach is well established and has enabled quantification of the 2D disorder potential \cite{Burgess2013}.  For a pristine structure in the vortex state, the transverse bias shifts sideways the path followed by the core during the $H_x$ field sweep, by an amount dictated by the displacement susceptibility.  In the presence of disorder, the characteristic $H_y^{DC}$ field scale for changing the Barkhausen `fingerprint' can be related to the characteristic length scale of the microstructure-induced local variation in magnetic properties.  More generally, this field scale can be determined by a convolution of the disorder with the high energy density spin texture element it interacts with.

Figure 5 shows select traces from a detailed mapping of the $z$-torque hysteresis across successive increments of $H_y^{DC}$.  As a function of $H_y$, most disorder-induced features in the data emerge and vanish over a span of 500$\thinspace$A/m or less.  This again is consistent with a characteristic length scale for the disorder in the range of 10$\thinspace$nm.  The low field core position versus field as described by the rigid vortex model \cite{Guslienko2001,Burgess2010} yields estimates for displacement susceptibilities, in a crude approximation for the double vortex state, of $\sim 30\thinspace$pm/(A/m) for the left-hand side and $\sim 20\thinspace$pm/(A/m) for the right-hand side core.  Also clearly visible in Fig.$\thinspace$5 are tunings with $H_y$ of the core nucleation fields for both transitions (single and double vortex states) in the field sweep down hysteresis branch.

\begin{figure}[ht] 
	\centering
	\includegraphics[scale=1] {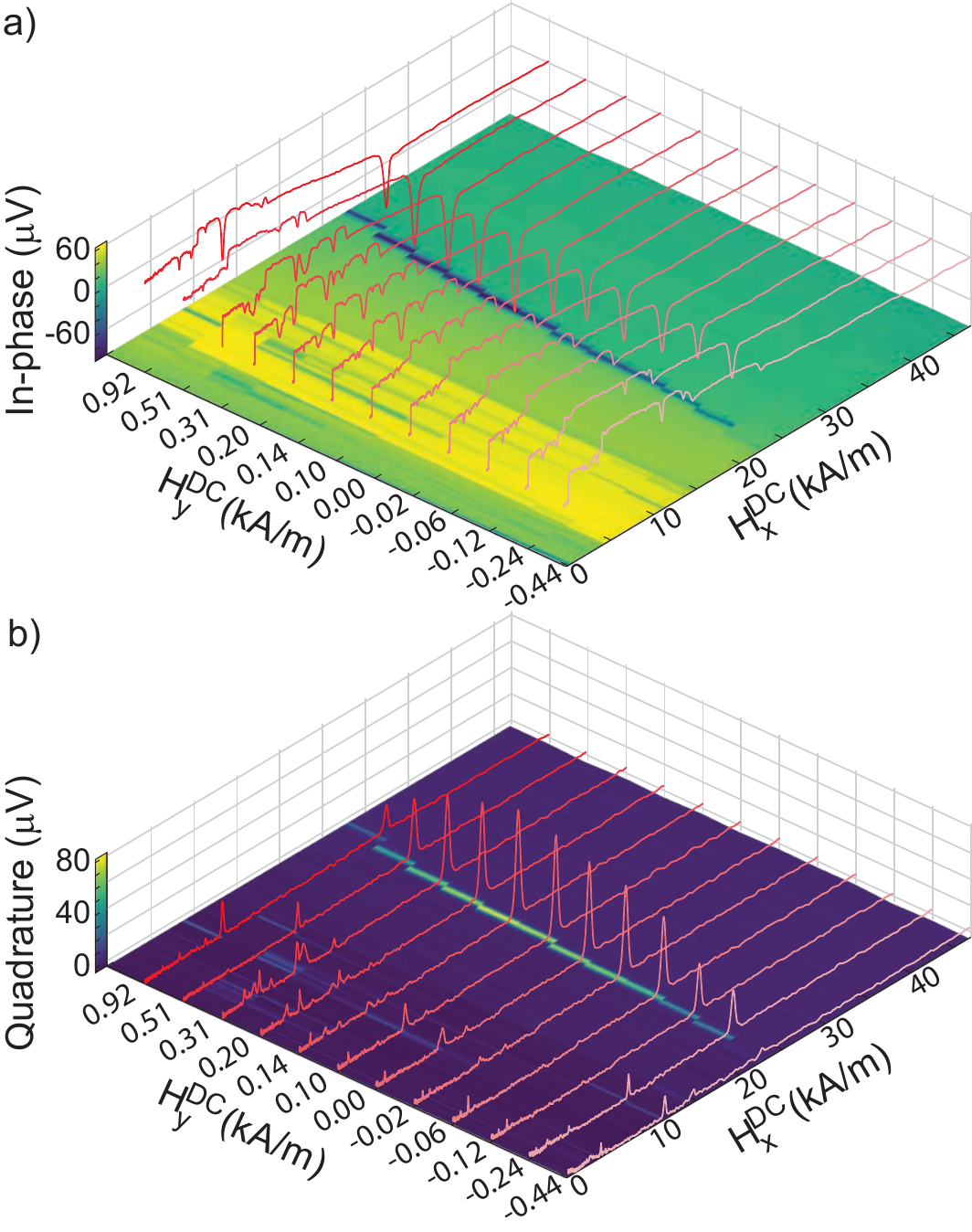}
	\caption{Experimental measurement of the 8.02$\thinspace$MHz ($\tau_z$) mode response to lowering $H_x$ from high field, for various values of DC field in the $y$-direction. a) In-phase component of the measured signal. b) Quadrature component of the measured signal. Only the high-to-low branch is shown.}
	\label{Fig5}
\end{figure}

Vortex cores are smaller than or comparable in size to individual pinning sites.  Domain walls, being so much larger than individual sites, are collectively pinned at multiple locations along their length, in any given configuration.  Only a small subset of collective pinning configurations give rise to an enhanced $y$-susceptibility, causing the domain wall-related features in the simulations to be even narrower in $H_x$ than those arising from a vortex core alone.  

Strong domain wall-driven enhancements in susceptibility persist over counter-intuitively large ranges of $H_y$, in some cases.  A dramatic example is visible in Fig.$\thinspace$5, where the peak of $\tau_z$ from Fig.$\thinspace$2b appears to tune continuously in $H_x$ position while $H_y$ is varied over a range of 2$\thinspace$kA/m.  Such tuning is somewhat reminiscent of spin resonance; it is possible, although highly unusual, for ferromagnetic spin textures to exhibit resonant spectral weight at a frequency as low as 8$\thinspace$MHz \cite{Saitoh2004}.  Fig.$\thinspace$5b shows the companion plot of the quadrature component of the signal for this data set. 

Significantly, the largest peak and a few of the others within this set of measurements exhibit the `smoking gun' of spin dynamics in the form of a continuous phase-shift of the response across the peak.  A complementary rendering of the $H_y^{\rm DC}=0$ data is shown in Fig.$\thinspace$6c, a polar plot of magnitude $R$ versus phase angle $\theta$ (equivalent to a plot of the quadrature component versus the in-phase signal).  Most of the data points in Fig.$\thinspace$6c fall close to $\theta =0$.  A small overall change of phase with $H_x$ is visible (and expected, for example as a result of a small shift in mechanical resonance frequency with field).  More significantly, the two largest peaks open up as loops in the polar plot, suggesting distorted `resonance circles'.  The continuous phase shifts and enhanced magnitude of response may arise from stochastic resonances of high energy density regions of the spin texture hopping between neighbouring pinning sites.  At these features, the periodic tilting of the energy landscape by the in-plane RF magnetic field yields a response magnified in amplitude by the thermal noise, or in other words, the nominally incoherent thermally-activated hopping is modulated sufficiently by the field to yield a coherent susceptibility contribution to the signal detectable by the lock-in.  This process also introduces phase shifts between the drive and the response, often in the range of 45$^{\circ}$ overall, although understanding the phase shift in detail can be challenging \cite{Gammaitoni1991, Dykman1992}.  A less likely explanation is that the measurement could be interrogating the low-frequency wing of a broad resonance mode that shifts with field (bearing in mind that the measurement is not frequency-swept, as in a conventional resonance measurement), hence also limiting the phase excursion. 

\begin{figure}[ht] 
	\centering
	\includegraphics[scale=1] {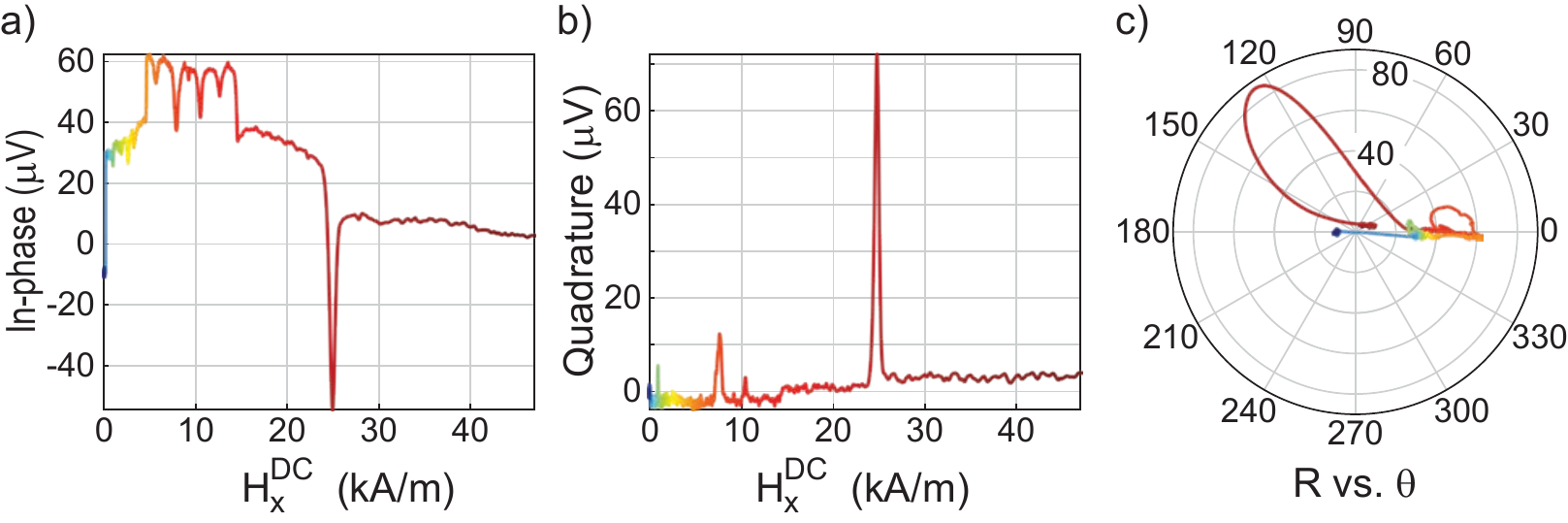}
	\caption{Magnetic response of the permalloy structure to the in-plane DC magnetic field in the $x$-direction with zero field in the $y$-direction for the 8$\thinspace$Mhz mode. a) In-phase component of the measured signal. b) Quadrature component of the measured signal. c) Polar plot of the measured signal. Only high-to-low branch is shown with colors (red to blue) indicating the proper direction in the polar plot.} 
	\label{Fig6}
\end{figure}

\section*{Summary}
The power and potential of optical nanocavity displacement sensing for multi-component physical measurements of condensed matter nanosystems has been demonstrated through a detailed study of mechanically-transduced in-plane and out-of-plane magnetic torques from a thin-film permalloy microstructure.  The torques arise both from the nonuniform arrangement of magnetization within the structure (the spin texture), as dictated by the shape, size, and external field strength and direction, and also from the interaction of high energy density regions of the texture with weak magnetic disorder, overall providing a very detailed picture.  The global and local magnetic anisotropies yield RF torques sensing the curvature of the total magnetic energy versus applied field direction.  Very importantly for thin-film structures, the out-of-plane torque is much more sensitive to the `springiness' of the spin texture, including the exchange energy density contribution.  This opens new opportunities for coupling spin dynamics to the cavity optomechanics.  It will also be a valuable tool in the study of exchange bias and other interface effects \cite{daSilva2018}.  

Remarkably, the ability of single optical nanocavities to detect orthogonal directions of displacement can be just as important a feature as the cavity-enhanced sensitivity, in recommending their use.  This feature should be extensible also to vectorial force measurements \cite{Rossi2017, Rossi2019}.  Inspired by the force studies, where Rabi oscillations of coupled flexural modes have been observed \cite{Braakman2018}, magnetically-mediated coupling between mechanical torsion modes may be contemplated.     

\section*{Acknowledgements}
GH is supported by an NSERC Postdoctoral Fellowship award.  We are grateful to Michael Dunsmore for the COMSOL simulations to estimate degrees of torsionality of the mechanical modes, and to Jonathan Leliaert for guidance in configuring the mumax$^3$ simulations at nonzero temperature.

\section*{References}

\bibliographystyle{abbrv}
\bibliography{refs-v7-09aug2019}



\pagestyle{headings}


\newpage

\end{document}